\documentstyle[aps,preprint]{revtex}

\newcommand{\be}{\begin{equation}}
\newcommand{\ee}{\end{equation}}
\newcommand{\bea}{\begin{eqnarray}}
\newcommand{\eea}{\end{eqnarray}}

\begin{document}
\draft
\title{Gutenberg Richter and Characteristic Earthquake 
behavior in Simple Mean-Field Models of Heterogeneous Faults}
\author{Karin Dahmen and Deniz Erta{\c{s}\cite{denizinfo}}}
\address{Lyman Laboratory of Physics, Harvard University, Cambridge, 
Massachusetts, 02138}
\author{Yehuda Ben-Zion}
\address{Department of Earth Sciences, Univ. of Southern CA,
Los Angeles, CA, 90089-0740}

\date{\today}

\maketitle

\begin{abstract}
The statistics of earthquakes in a heterogeneous fault zone is studied
analytically and numerically in a mean field version of a model
for a segmented fault system in a three-dimensional elastic 
solid\cite{Yehuda,Fisher}. The studies focus on the interplay
between the roles of disorder, dynamical effects, and driving mechanisms. A
two-parameter phase diagram is found, spanned by the amplitude of dynamical
weakening (or ``overshoot'') effects $\epsilon $ and the normal distance $L$
of the driving forces from the fault. In general, small $\epsilon $ and
small $L$ are found to produce Gutenberg-Richter type power law statistics
with an exponential cutoff, while large $\epsilon $ and large $L$ lead to a
distribution of small events combined with characteristic system-size
events. In a certain parameter regime the behavior is bistable, with 
transitions back and forth from one phase to the other
on time scales determined by the fault size and other model
parameters. The implications for realistic earthquake statistics are
discussed.
\end{abstract}




\narrowtext

\section{Introduction}
The statistics of earthquakes has been a subject of 
research for a long time. One spectacular feature is the wide range 
of observed earthquake sizes, spanning over ten 
decades in earthquake moment magnitude (which is defined to scale as the
logarithm of the integral of slip along the fault during the 
earthquake\cite{Gutenberg}). 
Gutenberg and Richter\cite{Gutenberg} found
in the 50's that the size distribution of regional earthquakes follows
a power law over the entire range of observed events. 
The exponent $b$ of the power law distribution appears to be
universal, {\it i.e.} it is approximately the same 
(within statistical errors and possible secondary dependency
on the tectonic domain) for all studied regions. This type of power law
distribution is called the ``Gutenberg Richter'' distribution.
Recently, enough data has been collected to 
extract statistics on individual systems of earthquake faults,
or more precisely on systems of narrow fault zones.
Interestingly, it was found that 
the distribution of earthquake magnitudes may vary
substantially from one fault system to another. 
In particular, Wesnousky and coworkers\cite{Wesnousky} found that 
fault systems with highly irregular geometry, such as the San 
Jacinto fault zone in California,
which have many offsets and branches, display
``power law'' statistics over the whole range of observed
magnitudes. 
Not all fault systems, however, display a power law
distribution on all scales up to the largest earthquakes. 
The available data\cite{Wesnousky} 
indicate that fault systems with more regular geometry
(presumably generated progressively with increasing cumulative slip)
such as the San
Andreas fault in California display power law distributions only for small
events, which occur in the time intervals between roughly quasi-periodic 
earthquakes 
of a much larger ``characteristic'' size which rupture the
entire fault. There are practically no observed earthquakes of intermediate
magnitudes on such geometrically regular fault systems.
Distributions of this type are called the ``characteristic earthquake'' 
distribution.

In previous work\cite{Yehuda,Fisher} it was demonstrated that a class of
simple models of ruptures along a heterogeneous fault zone
displays both types of behavior. 
The universal power law scaling behavior of earthquake statistics was seen
to be due to an underlying critical point, which becomes mean-field-like
for fault geometries with more than two spatial dimensions. In the 
limit of {\it weak} dynamical effects, the mean-field
approximation to the 2 dimensional fault provides a more appropriate 
approximation than, for example, traditionally studied one dimensional 
approximations to the models. In fact, {\it exact} results 
for the scaling exponents (up to logarithmic corrections) could be
obtained from mean-field theory. The reason is that 
the elastic stresses along the fault are effectively long range (decaying
like the inverse cube of the distance), 
such that in two and higher dimensions
the fluctuations due to interaction with other points on the fault decrease
as the fault size is increased -- on long length scales the behavior becomes
the same as that of a system with infinite ranged elastic interactions 
(up to logarithmic corrections in two dimensions). In other words,
the upper critical
dimension is equal to the physical dimension of the fault, which
is 2\cite{Obukhov,Deniz}. (Some of the static mean-field exponents 
turned out to be the same as in other quasi-static models\cite{Obukhov}.)
In the presence of small but nonzero weakening effects of amplitude $\epsilon$
a critical rupture size (slipping area) $n_{cr}$ for ``runaway'' or 
``characteristic fault size'' events was calculated perturbatively
\cite{Fisher} and was 
found to scale as $1/\epsilon^2$. Faults of larger area than this size are 
expected to display the characteristic earthquake distribution, with small 
events up to size $n_{cr}$, and no events of intermediate
size between $n_{cr}$ and the characteristic fault size events.
For faults of smaller total area than $n_{cr}$ only the power law scaling
region of the small events is seen, so the distribution is of the 
Gutenberg Richter type.

In this paper we examine a mean-field model with a range of 
dynamical weakening effects from weak to strong, 
and different levels of disorder in the 
brittle properties. Specifically, we study the model of Ben-Zion 
and Rice\cite{Yehuda}, which involves simple approximations of 
dynamic frictional weakening (similar to static versus dynamic friction),
but replace the physical long range elastic interactions with infinite
range interactions. In addition to exhibiting both ``power law'' and
``characteristic'' scaling of event sizes, this model exhibits the
possibility of {\it coexistence }of these two types of behavior. That is, for a
given set of model parameters, the system has {\it two distinct persistent
stationary states}. In an infinitely large system it will depend on the
initial conditions whether the system displays Gutenberg Richter
or characteristic earthquake type behavior.
Faults of {\it finite} size can spontaneously switch from one state
to the other on time scales that are exponentially large in system size.
The switching times (or ``persistence times'') are determined by nucleation 
processes from
one state to the other, similar to flips back and forth at coexistence
in finite thermally equilibrated systems.
Many of the
qualitative features seem to be sufficiently robust to be applicable to real
fault zones. Interesting to note, such ``switching'' behavior appears to 
characterize
long paleoseismic records observed along the Dead Sea transform fault system
in Israel\cite{Israelx1}, and is compatible with other paleoseismic
\cite{Californiax2} and geologic\cite{westernUSx3} data. 
In addition, qualitatively similar 
switching has been recently found in regional models of disordered
fault systems\cite{yehudax3}.

The remainder of this paper is organized as follows: In Section \ref
{secmodel}, we define the model and provide a summary of the main results.
In Section \ref{secanalysis}, we present a detailed analysis of the model
along with comparisons with numerical simulations. 
In Section \ref{secdiscuss}, we compare our results with earlier studies 
of similar models
and discuss their potential relevance to natural fault systems modeled as a 
narrow fault zone in a three dimensional elastic surrounding media.

\section{The Model and Summary of Results}

\label{secmodel}

Ben-Zion and Rice\cite{Yehuda} suggested that a heterogeneous fault system with
offsets and branches may be represented by an array of discrete cells in
a two dimensional plane,
with spatially varying ``macroscopic'' constitutive parameters that model
the heterogeneity of the original fault system. This model fault on the $%
(x,z)$ plane can be considered as a collection of brittle patches mapped 
onto the interface between two tectonic blocks, which move with (small) 
relative transverse velocity $v\widehat{{\bf x}}$ far away from the fault. 
In the simple realizations used in refs.\cite{Yehuda,Fisher}, and here,
(as in related models \cite{Knoppoff}),
the fault plane is segmented into $N$ geometrically equal cells. 
In the mean field approximation of infinite range elastic interactions,
the local stress $\tau _{i}$ on cell $i$ is
given by

\begin{eqnarray}
\tau _{i} &=&J/N\sum_{j}(u_{j}-u_{i})+K_{L}(vt-u_{i})  \nonumber \\
&=&J{\bar u} +K_{L}vt-(K_{L}+J)u_{i},  \label{eqstress}
\end{eqnarray}
where $u_{i}$ is the total fault offset of cell $i$ in the horizontal $(x)$
direction, ${\bar u} =(\sum_{j}u_{j})/N$, $J/N$ is the elastic
coupling between cells in the mean-field approximation, and $K_{L}$ is the
effective loading stiffness of the bulk material surrounding the fault patch.

Initially, the fault is in a relaxed configuration, i.e. all stresses are
less than a local {\it static} failure threshold stress $\tau _{s,i}$. 
In the absence of brittle failures the stresses at the cells increase
uniformly due to the external loading and $\dot{\tau}_{i}=K_{L}v$.
As long as no cell reaches its failure threshold, $\dot{u}_{i}=0$ 
everywhere. When the stress at a cell becomes larger than $\tau
_{s,i}$, the cell slips by an amount $\delta u_{i}=(\tau _{s,i}-\tau
_{a,i})/(K_{L}+J)$, to reduce its stress from $\tau _{s,i}$ to an arrest
stress $\tau _{a,i}$. (The nonuniformity of failure and arrest stresses
across the fault plane models the spatial heterogeneity of real fault zones%
\cite{Yehuda}.) Consequently, during failure
cell stresses change by [cf. Eq.(\ref{eqstress}%
) ] 
\begin{mathletters}
\label{eqfailure}
\begin{eqnarray}
\delta \tau _{i} &=&\tau _{a,i}-\tau _{s,i},  \\
\delta \tau _{j} &=&(c/N)(\tau _{s,i}-\tau _{a,i}),\qquad j\neq i,
\end{eqnarray}
where $c\equiv J/(K_{L}+J)$ is a ``conservation parameter'' giving 
the fraction of the stress drop of the failing cell retained in the 
system after the slip. As
pointed out in Refs.\cite{Fisher,KLzero}, for fault zones with characteristic linear
dimensions of $O(L)$, the ``loading spring constant'' is $K_{L}\sim 1/L,$
provided that the stress loading of the fault is either due to uniformly
moving (creeping) boundaries or applied forces at distances of $O(L)$ 
away
from the fault plane. For the case $N=L^2$,
$(1-c)\sim O\left( 1/\sqrt{N}\right) .$
A value $c<1$ for a large system would be physically realized if the
external drive is closer to the fault than its linear extent. 

During the failure process, the slipped cell is assumed to be weakened by the
rupture, such that its failure strength is reduced to a {\it dynamical}
value $\tau _{d,i}\equiv \tau _{s,i}-\epsilon (\tau _{s,i}-\tau _{a,i})$,
with $0\leq \epsilon \leq 1$ parameterizing the relative importance of
dynamical weakening effects in the system. If the failure stress
transfer brings other cells to their failure threshold, an avalanche of
cell failures, i.e., ``rupture propagation'', 
occurs according to Eqs.(\ref{eqfailure})
until all cells are at stresses $\tau _{i}\leq \tau _{s,i}$\cite
{explanation}. It is assumed that these avalanches happen on time scales
short compared to the external loading time (i.e. $v$ is adiabatically 
small), so that
the external load is kept constant during an earthquake. In time intervals
between
earthquakes, all cells are assumed to heal completely, thus failure
thresholds are reset to their static value $\tau _{s,i}$ and the external
loading resumes until the next cell failure.

In order to simplify notation, it is useful to introduce rescaled stress 
variables

\end{mathletters}
\begin{eqnarray}
s_{i} &\equiv &1-\frac{\tau _{s,i}-\tau _{i}}{\langle \tau _{s,i}-\tau
_{a,i}\rangle }, \\
s_{a,i} &\equiv &1-\frac{\tau _{s,i}-\tau _{a,i}}{\langle \tau _{s,i}-\tau
_{a,i}\rangle }, \\
s_{d,i} &\equiv &1-\frac{\tau _{s,i}-\tau _{d,i}}{\langle \tau _{s,i}-\tau
_{a,i}\rangle }=1-\epsilon (1-s_{a,i}),
\end{eqnarray}
such that cell failure always occurs when $s_{i}=1,$ and $\langle
s_{a}\rangle =0$. (Here, $\langle \rangle$ symbolizes averaging over all
cells in the fault zone.)
The arrest stress $s_{a,i}$ is uncorrelated from cell to cell, and
is picked once for each segment 
from a probability distribution $\rho (s_{a})$ with mean $0$ and a
compact support $(-W/2,W/2)$ of width $0\leq W\leq 2$, which characterizes
the heterogeneity of the fault system. [In our simulations we have used the
parabolic distribution $\rho (s_{a})=3(W^{2}-4s_{a}^{2})/(2W^{3})$, for $%
-W/2\leq s_{a}\leq W/2$ and $0$ otherwise.] Unless stated otherwise, the
focus is on the small disorder limit $W$ $\ll 1$ and moderate values for $%
\epsilon $, which are considered fixed, and the properties of the system are
analyzed as a function of varying conservation parameter $c$ and system size 
$N$. (In the last section of the paper we discuss the effects of 
larger values of $W$ as well.)
The size of an earthquake refers to the number of cells that failed,
(i.e. the ``area'' on the fault that slips in an earthquake).

For $N \rightarrow \infty$,
depending on relative values of the system parameters, there are in general
two possible steady-state distributions of cell stresses and of earthquake
magnitudes. We refer to these as
as ``phases'':

(A) {\it The ``Gutenberg-Richter'' (G-R) Phase: } This phase, possible
in both regions $1$ and $2$ of Fig.~\ref{phasediag},
is characterized by a distribution of earthquake sizes $p_{e}^{(f)}(n)$
of power law form. In infinite systems ($N \rightarrow \infty $),
it is given by
\begin{equation}
p_{e}^{(f)}(n)\approx A_{f}n^{-3/2}\exp (-n/n_{cf}),\qquad n\ll N,
\label{eqpef}
\end{equation}
with a characteristic cutoff size $n_{cf}\approx 2(1-c)^{-2}$ that diverges
as $c\nearrow 1$. 
(Finite size corrections are given in equation (\ref{eqpefc}) below).
The stress $%
s_{i}$ at a given cell is independent of all others and is equally likely to
take any allowable value, i.e., 
\begin{equation}
\text{Prob}(s\leq s_{i}\leq s+ds)=\frac{ds}{1-s_{a,i}},\qquad s_{a,i}\leq
s\leq 1.
\end{equation}
Thus, the stress distribution in the G-R phase is given by 
\begin{equation}
p^{(f)}(s)=\int_{-\infty }^{s}ds_{a}\frac{\rho (s_{a})}{1-s_{a}},
\label{eqpf}
\end{equation}
which is uniform and equal to $\bar{p}\equiv \langle (1-s_{a})^{-1}\rangle $
in the interval $(W/2)\leq s\leq 1$ [see Fig.~\ref{figpe}(a), inset].
In this phase
each cell fails at most once during an earthquake, and therefore dynamical
effects are largely irrelevant. 
An infinitely large system which started in the 
G-R phase will remain it this phase. In finite systems $N < \infty $,
with parameters in region $2$ of figure \ref{phasediag}, however,
a very large earthquake of size $(1-\epsilon)N/c$
or greater occasionally 
triggers dynamical effects that lead to a catastrophic
``runaway'' event in which all cells eventually fail and cause a substantial
change in the stress distribution and subsequent evolution of the system, as
outlined next.

(B) {\it The ``Runaway'' Phase:} This phase is characterized by a
quasi-periodic occurrence of system wide earthquakes in which all cells
fail. As a result of dynamical effects, the stress $s_{i}$ in a cell
immediately after such a ``runaway'' event is independent of other cells and
is equally likely to take any value between its arrest stress and dynamical
failure stress, i.e., 
\begin{equation}
\text{Prob}(s\leq s_{i}\leq s+ds)=\frac{ds}{s_{d,i}-s_{a,i}},\qquad
s_{a,i}\leq s_{i}\leq s_{d,i}.  \label{eqrd}
\end{equation}
The stress distribution is thus given by 
\begin{equation}
p^{(r)}(s)=\frac{1}{1-\epsilon }\int_{\frac{s-(1-\epsilon )}{\epsilon }%
}^{s}ds_{a}\frac{\rho (s_{a})}{1-s_{a}},  \label{eqpr}
\end{equation}
which is uniform and equal to $\bar{p}/(1-\epsilon )$ in the interval $%
(W/2)\leq s\leq 1-\epsilon -(\epsilon W/2)$ [see Fig.~\ref{figpe}(b),
inset]. The runaway event is followed
by a quiescent period during which stresses on the cells build back up from
their dynamic failure value to near their static failure value. Subsequent
small events are followed by the next runaway event, at which
point the stress distribution is reset to Eq.(\ref{eqpr}). 
These background small events have a
size distribution similar to events in the G-R phase, but with a different
cutoff size. In an infinite system (for finite size corrections see
equation (\ref{eqperc}) below):
\begin{eqnarray}
p_{e}^{(r)}(n)\ &\approx &A_{r}n^{-3/2}\exp (-n/n_{cr}),\qquad n\ll N,
\label{eqper} \\
n_{cr} &\approx &\frac{2(1-\epsilon )^{2}}{(1-\epsilon -c)^{2}},
\end{eqnarray}
which diverges as $c \searrow (1-\epsilon )$. However, this divergence is
never observed, as the runaway phase becomes unstable against breakup into
the G-R phase when $c<c^{*}\equiv (1+\epsilon )^{-1}$ for the following
reason: If the background small events during a cycle involve at least a 
fraction $r_{c}$
of the cells, the subsequent large event is unable to cause all of the
cells to fail, since the cells that failed during background 
activity are farther
away from their failure stress. This typically causes a spontaneous breakup
of the bunched stress distribution and a resumption of the G-R phase. The
fraction of cells needed to cause this breakup is given by $r_{c}=1+\epsilon
-c^{-1}=(c^{*})^{-1}-c^{-1}$ as is derived in section \ref{secanalysis}.
When $c\searrow c^{*},$ the size of 
background events
necessary to cause breakup vanishes and the runaway phase becomes unstable,
{\it i.e.} for $c < c^{*}$ the G-R phase is the only persistent 
phase, regardless of the initial conditions.
For $c> c^{*}$, the G-R phase and the Runaway phase are both persistent in
an infinite system. In an infinite system
the initial conditions determine which one the system
displays.
In a finite system, however, exponentially rare earthquakes can lead to
nucleation from the G-R phase into the Runaway phase and vice versa.
Equations (\ref{timefr}) and (\ref{timerf}) below give estimates of the times 
spent in the respective phases between such nucleation (or ``switching'') 
events.

\section{Analysis of the model}

\label{secanalysis}The results quoted above have been obtained by mapping
earthquakes in the model to corresponding events in a stochastic process,
which is approximated by a series of Bernoulli trials\cite{Feller} in order
to be able to obtain analytical estimates for the various quantities of
interest, such as distributions of earthquake sizes and persistence times
for the two phases.

\subsection{Gutenberg-Richter Phase}

Let us first focus on the G-R phase. At some instant $t$ immediately
preceding a cell failure, consider the sequence $\left\{ X_{n}\equiv
1-s_{i(n+1)}\right\} ,$ where $i(n)$ is the index of the cell that has the $%
n $th largest stress in the system [See Fig.~\ref{figpoisson}]. For a large
system, the stress gaps $\left\{ \delta s_{n}=X_{n-1}-X_{n}\right\} $ are
(almost) independent of each other, drawn from an exponential probability
distribution, i.e., Prob$\left( \delta s_{n}=s\right) =\exp \left( -\bar{p}%
Ns\right) $, with  $\bar{p}\equiv \langle (1-s_{a})^{-1}\rangle $.
For $n\gg 1,\ X_{n}$ resembles a biased random walk with a mean 
$\mu _{X}(n)=n/(\bar{p}N)$ and variance $\sigma _{X}^{2}(n)=n/(\bar{p}N)^{2}$%
. As long as dynamical effects are absent, the stress redistributed to each
cell following the failure of the first $n$ cells is given by a random
variable $Y_{n}$ with mean $\mu _{Y}(n)=nc/(\bar{p}N)$ and variance $\sigma
_{Y}^{2}(n)\approx n(cW/\bar{p}N)^{2}\ll \sigma _{X}^{2}(n).$ A triggered
earthquake can sustain itself only if the redistributed stresses exceed
the stress gaps. Therefore, $Z_{n}\equiv X_{n}-Y_{n}<0$ during an earthquake
and it immediately follows that the distribution of earthquake sizes for $%
N\gg n\gg 1$ are given in terms of the distribution of first passage times of
biased random walks. Approximating the continuous probability distribution
of the step sizes of $\{Z_{n}\}$ with a Bernoulli process (where steps of
equal size are taken up or down with probability $p$ and $1-p,$
respectively), we can utilize results available for Bernoulli trials\cite
{Feller}, 
\begin{equation}
\text{Prob}\left( Z_{i}<0,0<i<n;Z_{n}=0\right) =\frac{\text{Prob}(Z_{n}=0)}{n%
},
\end{equation}
i.e. the probability for the {\it first} return to the origin after $n$
steps equals the total probability of reaching the origin after $n$ steps
divided by $n$. 
$Prob[Z_n = 0]$ can easily be calculated\cite{Feller} for 
$N \gg n\gg 1$. One obtains Eq.(\ref{eqpef}) with $n_{cf}=2(\mu
^{2}+\sigma ^{2})/\mu ^{2},$ where $\mu $ and $\sigma ^{2}$ are the mean and
variance of the step size for the process $Z_{n},$ respectively.
Substituting the values $\mu =(1-c)/(\bar{p}N)$ and $\sigma ^{2}\approx 1/(%
\bar{p}N)^{2},$ the cutoff length is given by 
\begin{equation}
n_{cf}\approx 2(1+(1-c)^{-2})=\frac{2}{\left( 1-c\right) ^{2}}\left\{
1+O\left[ (1-c)^{2}\right] \right\} ,  \label{eqncf}
\end{equation}
where the last approximation is justified since treating $Z_{n}$ as a
Bernoulli process is expected to yield relative errors of $O(\mu ^{2}/\sigma
^{2}).$ 

For finite-sized systems, when the fraction of failed cells $r=n/N$
is no longer small, Eq.(\ref{eqpef}) needs to be modified since the stress
gaps are not entirely independent: In order to correctly reflect the fact
that $Z_{N}=1-c$ to within $O(1/N),$ the Bernoulli process should be
constrained to return to its mean value after $N$ steps. This can be
achieved by calculating the corresponding conditional probabilities: 
\begin{eqnarray}
p_{e}^{(f)}(n) &=&\frac{\text{Prob}(Z_{n}=0)}{n}\frac{\text{Prob}%
(Z_{N-n}=1-c)}{\text{Prob}(Z_{N}=1-c)}  \nonumber \\
&=&\frac{\tilde{A}_{f}}{n^{3/2}}\exp \left\{ -\frac{n(1+n/N)}{n_{cf}}%
\right\} ,  \label{eqpefc}
\end{eqnarray}
which reduces to Eq.(\ref{eqpef}) in the limit $n\ll N.$ ( $\tilde{A}_{f}$
is a constant fixed by normalization.) Figure \ref{figpe}(a) shows the
distribution of event sizes for numerical simulations of the model with $%
N=400,$ for values $c=0.6,0.7$ and $0.8.$ (In all presented simulation
results, $W=2/19$ and $\epsilon =0.5.)$ The continuous lines are
one-parameter fits to the form (\ref{eqpefc}). The discrepancy between the
fitted and theoretical [from Eq.(\ref{eqncf})] values of $n_{cf}$ is
consistent with the expected relative error.

As mentioned earlier, the failure of all the remaining cells becomes
very likely once $(1-\epsilon )N/c$ cells have failed, since the initially
failed cells reach their dynamical failure stress. The mean event size 
is roughly equal to $n_{cf}^{1/2}$, therefore the mean time between events
is $T_0 n_{cf}^{1/2}/N$, where $T_{0}\equiv \langle \tau _{s,i}-\tau _{a,i}
\rangle /(K_{L}v)$ is the characteristic time over which a cell is loaded 
from its arrest stress to its failure stress. The typical waiting
time to see a switch to the runaway phase yields [cf. Eq.(\ref{eqpefc})] 
\begin{equation}
\label{timefr}
T_{f\rightarrow r}\approx T_{0}\frac{C_{fr}N^{1/2}}{n_{cf}^{1/2}}\exp \left\{ 
\frac{(1-\epsilon )(1-\epsilon +c)}{c^{2}n_{cf}}N\right\} ,
\end{equation}
with $C_{fr}$ a factor of order unity which varies weakly
with $\epsilon$ and $c$ in the region of interest, provided that each
attempt is statistically independent of each other. 
We verified that in our simulations indeed no time correlations
of event sizes were present, out to many times the characteristic time
$T_0$ (see also the discussion section).
The distribution of
persistence times should then obey Poisson statistics with mean $%
T_{f\rightarrow r}$. Figure \ref{figswitchtimes} depicts the distribution of
persistence times (with a fit to Poisson statistics) for $N=400$ and $c=0.73$.
Mean persistence times depend very sensitively on the conservation parameter $%
c,$ as shown in the inset of Fig. \ref{figswitchtimes}\cite{footchannels}.

\subsection{Runaway Phase}

Let us next consider the runaway phase. Immediately preceding the first cell
failure after a runaway event, the stress gaps $\left\{ \delta
s_{n}=X_{n-1}-X_{n}\right\} $ have the probability distribution Prob$\left(
\delta s_{n}=s\right) =\exp \left[ -\bar{p}Ns/(1-\epsilon )\right] .$ Hence, 
$\{X_{n}\}$ has a mean $\mu _{X}(n)=n(1-\epsilon )/(\bar{p}N)$ and variance $%
\sigma _{X}^{2}(n)=n(1-\epsilon )^{2}/(\bar{p}N)^{2}$. As long as dynamical
effects are absent, the stress redistributed to each cell following the
failure of the first $n$ cells is still given by $\{Y_{n}\}$ with mean $\mu
_{Y}(n)=nc/(\bar{p}N)$ and variance $\sigma _{Y}^{2}(n)\approx n(cW/\bar{p}%
N)^{2}\ll \sigma _{X}^{2}(n).$ Thus, the mean and variance of the step size
for $\{Z_{n}\}$ is $\mu =(1-\epsilon -c)/(\bar{p}N)$ and $\sigma ^{2}\approx
(1-\epsilon )^{2}/(\bar{p}N)^{2}.$ The probability for an earthquake to
terminate after $n$ cell failures is [including finite size corrections
in analogy with Eq.(\ref{eqpefc})]
\begin{eqnarray}
p_{e}^{(r)}(n) &\approx &\frac{\tilde{A}_{r}}{n^{3/2}}\exp \left\{ -\frac{%
n(1+n/N)}{n_{cr}}\right\} ,\qquad \\
n_{cr} &=&\frac{2\left( 1-\epsilon \right) ^{2}}{\left[ c-(1-\epsilon
)\right] ^{2}}\left\{ 1+O\left( \left[ c-(1-\epsilon )\right] ^{2}\right)
\right\} ,\qquad c>1-\epsilon .
\label{eqperc}
\end{eqnarray}
Since $\mu <0$ for $1-\epsilon < c$, $Z_{n}<0$ with finite probability for
all $n$ and a runaway event occurs. In fact, a runaway event is inevitable
since $Z_{N}<0$ and the runaway event will commence once $\{Z_{n}\}$ reaches
its maximum. The total number of cells that fail before a runaway event is
given by the position of the maximum of $\{Z_{n}\}$, whose probability
distribution is proportional to $n_{cr}$ $p_{e}^{(r)}(n)$ for $n \gg n_{cr}$.
The mean number of these ``precursor'' cells is of order $n_{cr}^{1/2}$, 
which remains a finite constant as $N \rightarrow \infty$, {\it i.e.}, 
for big systems almost all the slip happens during the runaway 
events\cite{footprecursors}.

The remaining cells will all fail during the runaway event. Imagine a
situation where a fraction $r>(1-\epsilon )/c$ of the cells have failed. At
that point, the total redistributed stress per cell, is
\begin{eqnarray}
S &=&c\left[ r+\left( r-\frac{1-\epsilon }{c}\right) \left\{ c+c^{2}+\cdots
\right\} \right]   \nonumber \\
&=&\frac{c[r-(1-\epsilon )]}{1-c},
\end{eqnarray}
where the second term arises from repeat failures of some cells. $S\geq 1$
is needed to ensure that small event cells fail again and recreate the stress
distribution (\ref{eqpr}). This is achieved if 
\[
r\geq r^{*}\equiv \frac{1}{c}-\epsilon .
\]
Thus, the large event cannot recreate the stress distribution (\ref{eqpr})
if more than $(1-r^{*})N=r_{c}N$ cells fail during background activity. 
This usually leads to
a breakup of the bunched stress distribution and subsequent evolution
towards the G-R stress distribution (\ref{eqpf}). The typical persistence time
of the Runaway phase before a switch to the G-R phase is \cite{footchannels}
\begin{equation}
\label{timerf}
T_{r\rightarrow f}\approx T_0 \frac{C_{rf}N^{3/2}}{n_{cr}^2}
\exp \left\{ \frac{(c-c^{*})[1+(c-c^{*})/(c^{*}c)]N}{c^{*}c\,n_{cr}%
}\right\} ,\qquad c>c^{*},
\end{equation}
provided that all attempts are statistically independent of each other. 
(We have explicitly checked in the simulations that in
the runaway phase, the particular realizations of stress distributions
immediately following a large event are statistically independent of each
other\cite{footbreakup}). $ T_{r\rightarrow f}$ becomes comparable 
to the typical time between runaway events when 
$c \searrow c^{*},$ as expected\cite{ncrfoot}. Figure 
\ref{figswitchtimes} depicts the distribution of persistence times and a fit to
Poisson statistics for $N=100$ and $c=0.73$. The inset shows the dependence
of mean persistence times on $c$ for $N=100$. Although agreement with
Eqs.(\ref{timefr}) and (\ref{timerf}) is rather poor, the strong exponential 
dependence as a function of conservation parameter is evident.

\section{ Discussion}

\label{secdiscuss}The persistence times in both the G-R phase and the runaway
phase diverge exponentially with system size for $(1+\epsilon )^{-1}<c<1$,
and the system remains in either phase for extremely long times, thus the
phase space has two almost stable attractors. Clearly, the runaway phase 
represents
a more ``ordered'' stress distribution. Indeed, the basin of attraction for
the runaway phase is extremely small. In order to quantify this aspect,
consider the ``configurational entropy'' for a given stress distribution $p(%
\tilde{s}),$ with $\tilde{s}_{i}\equiv (\tau _{i}-\tau _{a,i})/(\tau
_{f,i}-\tau _{a,i}):$%
\begin{equation}
S_{conf}\left( \left\{ p\right\} \right) \equiv -\int d\tilde{s}\,p(\tilde{s}%
)\ln \left[ p(\tilde{s})\right] .
\end{equation}
For the G-R phase, $S_{conf}^{(f)}=0,$ indicating that a ``generic'' stress
distribution characterizes the G-R phase. On the other hand, in the 
runaway phase
\begin{equation}
S_{conf}^{(r)}=-\int d\tilde{s}\,p^{(r)}(\tilde{s})\ln \left[ p^{(r)}(\tilde{%
s})\right] =\ln (1-\epsilon ),
\end{equation}
indicating that the stress distribution is highly organized in that
phase. For discrete $N$ the stress distribution is approximated by a 
histogram of the stress values, and the integral is replaced by the sum 
over all bins of the histogram. 

The time evolution of the configurational entropy of the stress
distribution, calculated with a 10 bin histogram,
is depicted in Fig. \ref{figtimeseries} along with event sizes.
It is clear that $S_{conf}(t)$ can be used as an ``order parameter'', a
number that distinguishes the G-R phase and the runaway phase, with the
advantage that it can be determined at any instant. Histograms of event
sizes require a finite time interval to collect, and there is always the
danger of mixing events from one phase with the other, thereby confusing the
picture: The cumulative event size distribution over many persistence
times is a weighted average of two entirely different event distributions,
which obscures the underlying physical phenomena. $S_{conf}$ provides a
reliable way to separate the two phases and makes it possible to accumulate
accurate event size distributions for both of them.
Unfortunately, such a quantity cannot be determined from existing field data 
since the spatial distribution of stress is unknown.

So far, the discussion has centered around the $W\ll 1$ limit, and the
main role played by the heterogeneities has been the ``randomization'' of
the stress distribution at time scales over which all cells fail a few
times. The distribution of loading times,
over which the cells are loaded from the individual arrest stress to the
failure stress, has a mean $T_0 \equiv \langle \tau_{s,i} -\tau_{a,i}
\rangle/(K_l v)$ and standard deviation of order $W T_0$. Therefore,
the ``randomization" time, over which the stress variables $s_i$ become
roughly uncorrelated, is of order $T_0/W$. Thus, even for small 
$W$, for large enough $N$ this will be small compared  to the persistence
times, which scale exponentially in N [See Eqs.(\ref{timefr}) and 
(\ref{timerf})]. This ensures the consistency of the assumption that the
earthquakes are basically statistically independent of each other. The  
validity of this assumption of statistical independence can be explicitly
verified by examining the time correlations of event sizes numerically; 
indeed no trace of any correlation was found in our simulations, out to
many times the randomization time $T_{0}/W$. Likewise, we have explicitly  
checked that in the runaway phase, the particular realizations of stress
distributions immediately following a large event are statistically
independent of each other\cite{footbreakup}. 
 
For finite values of $W<2,$ we expect most of the
features to remain qualitatively unchanged: In the G-R phase, the
exponential cutoff size still diverges as $n_{cf}\sim (1-c)^{-2},$ and
although $c^{*}$ in general depends on $W$ and the shape of $\rho (s_{a}),$
there is still a persistent runaway phase for $c^{*}<c\leq 1.$ However, the
situation is likely to change qualitatively once arrest stresses can be
arbitrarily close to failure stresses, i.e. $W=2,$ {\it and} new values for
the arrest stresses are picked every time a cell fails\cite{footdisorder}.
This corresponds to the situation discussed in Ref.\cite{Fisher} for
finite-dimensional systems. Immediately upon introduction of dynamical
weakening $(\epsilon >0)$, $n_{cf}\sim \epsilon ^{-2}$ when $c$ approaches $%
1,$ i.e. the cutoff size no longer diverges. Furthermore, for $c=1$ the G-R
phase is no longer persistent since the persistence time $T_{f\rightarrow r}$
remains finite for large $N.$

Some of the results presented for the mean-field model, especially the
qualitative phase diagram, calculated exponents for the power law
earthquake distributions, and the divergence of the cutoff length scale, can
be expected to apply to models with realistic interactions, up to logarithmic
corrections. Such is because the underlying critical points that control these
exponents remain mean-field-like down to $2$ dimensional faults. This result
is firmly established for the $\epsilon =0$ case\cite{Fisher,Deniz}. At
finite $\epsilon $ one expects the nucleation size for the runaway phase,
which equals $(1-\epsilon )N/c$ in mean field theory, to become independent
of the system size, since elastic forces in the fault plane concentrate
stresses along the earthquake rupture front as the earthquake progresses.
Earthquakes bigger than a finite nucleation size $N_{crack}$ become
unstoppable in the presence of dynamic weakening effects and small disorder%
\cite{Yehuda}, and rupture the entire fault. Nevertheless, for $%
n_{cf}<N_{crack},$ the mean-field scaling results may still apply at finite $%
\epsilon $, provided that $W<2,$ i.e., there is a finite minimum stress drop
associated with each cell failure. For systems with $N_{crack}>N,$ this
range will extend all the way to the fault size. In this case, one
remarkable consequence is that since generically $\left( 1-c\right) \sim 1/%
\sqrt{N}$\cite{Fisher,Obukhov}, the cutoff size in the G-R phase $n_{cf}$ $%
\sim (1-c)^{-2}\sim N,$ i.e., earthquakes on individual 
fault zones obey power law statistics for
events up to a finite fraction of the entire system size. 

An important 
result is the possibility that a fault system might switch from a
``Gutenberg-Richter'' earthquake distribution spontaneously to a
``characteristic'' earthquake distribution, as in the mean-field model.
We note that calculations based on an entirely different model,
simulating the coupled evolution of regional earthquakes and
faults in a rheologically layered $3D $ solid\cite{yehudax3},
show similar behavior. Clear observation of such mode switching
in nature requires data sets spanning many thousands of years.
Paleoseismic studies attempt to construct long histories
of seismic events at given locations from sequences of 
displaces and highly disturbed rock layers.
Remarkably, the longest available paleoseismic records, documenting
large earthquake activity along the Dead sea transform in 
Israel\cite{Israelx1} 
appear to be characterized by alternating phases of intense 
seismic activity lasting a few thousands of years, and
periods of comparable length without large seismic events.
Other, qualitatively similar alternating deformation phases 
have been documented in the eastern CA shear zone\cite{Californiax2}
and the Great Basin Province in the western US\cite{westernUSx3}.

Another intriguing possibility might arise in a fault system
of weakly coupled segments driven under similar conditions.
The seismic response of such systems might exhibit a sort of 
``coexistence'', i.e., a fraction of the patches
might follow characteristic scaling whereas the others obey
Gutenberg-Richter scaling, giving rise to a hybrid event size distribution.
This may explain examples in the data of reference\cite{Wesnousky},
where the characteristic ``bump'' in the distribution was not very 
pronounced.
Finally, we note that part or all of the low magnitude seismicity
in the G-R phase may be too small to be detected by a seismic network.
In this case the spontaneous switching between the Runaway and G-R
phases may be interpreted as transitions from seismic response of a fault 
system to creep-like behavior.

\section{Acknowledgments}

We have greatly benefited from extensive discussions with
Daniel S. Fisher, Jim Rice, and Sharad Ramanathan.

KD gratefully acknowledges support from the Society of Fellows
of Harvard University, and NSF via DMR 9106237, 9630064, 
Harvard's MRSEC and Harvard's Milton Fund.
DE acknowledges support by the NSF through Grants No. DMR-9106237,
No. DMR-9417047, and No. DMR-9416910. YBZ was supported by the 
Southern California Earthquake Center (based on NSF Cooperative
Agreement EAR-8920136 and USFS Cooperative Agreement 14-08-0001-A0899).

\begin{figure}[tbp]
\caption{ {\it \ }Schematic phase diagram of the system. There is a
``coexistence'' of two persistent stationary states called Gutenberg-Richter
and Runaway phases, in a finite region of
parameter space, marked region ``(2) meta-stable''. For region $1$ given by
$c<c^*=1/(1+\epsilon)$
(line AB) one finds only small avalanches, i.e. the system is always
in the Gutenberg-Richter phase.
\label{phasediag}}
\end{figure}

\begin{figure}[tbp]
\caption{Histograms of event size distributions in the two stationary states
(phases), for $W=2/19,\epsilon =0.5,N=400.$ (a) The ``Gutenberg-Richter''
phase, characterized by a power-law earthquake distribution with an
exponential cutoff. Solid lines are fits to the analytic form
(\protect\ref{eqpefc})
with $n_{cf}$ as a fitting parameter. Also indicated are analytic estimates $%
n_{cf}^{(th)}.$ The inset shows a typical stress distribution of this phase
for $c=0.7$. The solid line is a fit to the analytic form 
(\protect\ref{eqpf}).
The nonuniform region near $s=0$ extends from $-W/2$ to $W/2$.
(b) The ``runaway'' phase, with a similar background
distribution and large characteristic events. The inset shows a typical
stress distribution for $c=0.8$. The solid line is a fit to the analytic 
form 
(\protect\ref{eqpr}).
The nonuniform region near $s=0$ extends from $-W/2$ to $W/2$. 
Near $s=\epsilon $ it extends from $\epsilon (1-W/2)$ to $\epsilon (1+W/2)$.
\label{figpe}}
\end{figure}

\begin{figure}[tbp]
\caption{The process $\{Z_{n}\}$, which shows the incremental amount of stress
needed to keep an earthquake going (see text for the precise definition). 
Each failure event corresponds to a segment of the process that starts out 
from a maximum up to that point and ends when it exceeds that level, and is 
marked as alternating circles and squares. The sample shown here,
which corresponds to the stress distribution shown in the inset of
Figure 2(a), 
depicts events of size $6,14,2,1,1,\cdots$.
The fault is loaded adiabatically between these events, during the intervals 
when $\{Z_{n}\}$ moves monotonically up from one maximum to the next. 
These are shown as dotted lines connecting consecutive events.
\label{figpoisson}}
\end{figure}

\begin{figure}[tbp]
\caption{Distribution of persistence times $T_{f\rightarrow r}$ and $%
T_{r\rightarrow f}$ for $W=2/19$, $\epsilon =0.5$, $N=100$, $c=0.73$. 
The lines are
fits of the cumulative probabilities to the Poisson distribution. 
(Simulations for systems with other parameters that allowed for many more
switches during the simulated times clearly also gave Poisson distributions
for the distribution of persistence times.)
Inset: The
dependence of persistence times on conservation parameter $c$, 
(triangles $T_{f\rightarrow r}$, circles $T_{r\rightarrow f}$)
for the same
values of $W,\epsilon ,$ and $N.$ Statistical errors are comparable to
symbol sizes.
\label{figswitchtimes}}
\end{figure}

\begin{figure}[tbp]
\caption{Sample time series of earthquake sizes (top), 
plotted together with the
conformational entropy $S_{conf}(t)$ (bottom) for 
$W=2/19,\epsilon =0.5,N=100,
c=0.73.$ (For the calculation of  $S_{conf}(t)$,
the simulated stress distribution was approximated by a 10 bin histogram 
of the stress values, which was evaluated immediately after each earthquake.)
The earthquake size distribution
changes drastically every time $S_{conf}$ toggles from $0$ to $\ln
(1-\epsilon )$, indicating a transition from one phase to the other.
A failed switching attempt from the G-R phase to the runaway phase
is seen at about $T/T_0 = 10900$.}

\label{figtimeseries}
\end{figure}


\begin{references}
\bibitem{explanation}  Note that cell stresses do not necessarily get
arrested at $\tau _{a,i}$ during an earthquake; which would have required
the slip amount to be $(\tau _{i}-\tau _{a,i})/(K_{L}+J)$ instead. We use
Eqs.(\ref{eqfailure}) in order to mimic continuous time. This choice also 
makes the model Abelian, i.e. whether a given
cell will fail during an earthquake does not depend on which order they
fail, simplifying the analysis.
\bibitem{footprecursors} For larger disorder $W$ there are yet another 
kind of background events. Their number is proportional to $N$.
In the regime considered here (small $W$, and moderate $N$), however,
those effects are negligible, as is also confirmed by our simulation
results.
\bibitem{ncrfoot}  Note that $n_{cr}$ remains finite when $c\searrow c^{*}$,
therefore the cutoff size for background small events never diverges; 
the underlying
``critical point'' is obscured by the breakup instability.
\bibitem{footbreakup}  One exception to the statistical independence of
events happens during the breakup of the runaway stress distribution, where
atypically large (for the G-R phase) earthquakes occur within a few $T_{0}$
of the initiation of breakup. These events should not be included in an
analysis of the event size distributions, since they are simply transients
and not representative of events in steady-state.
\end{references}
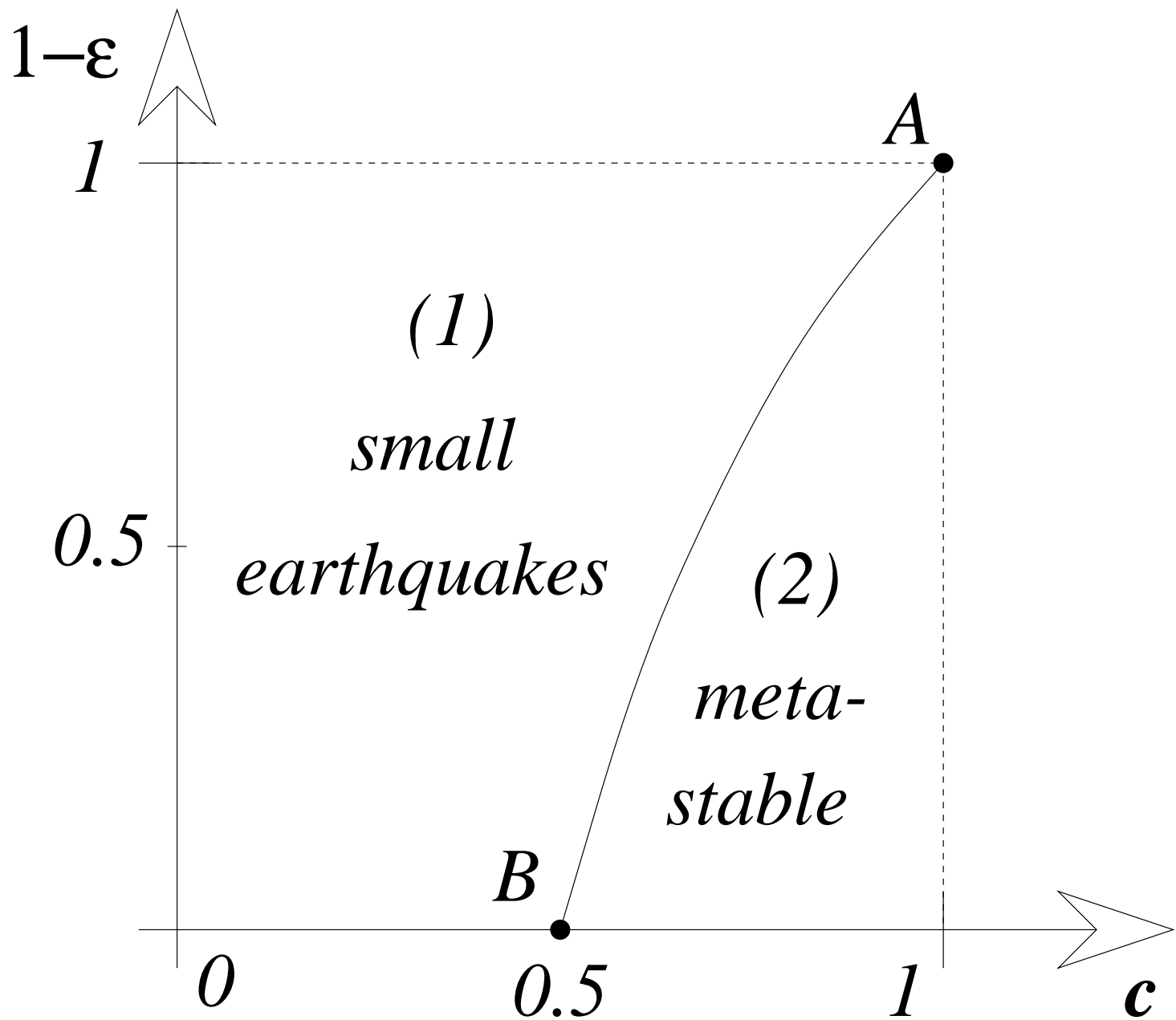
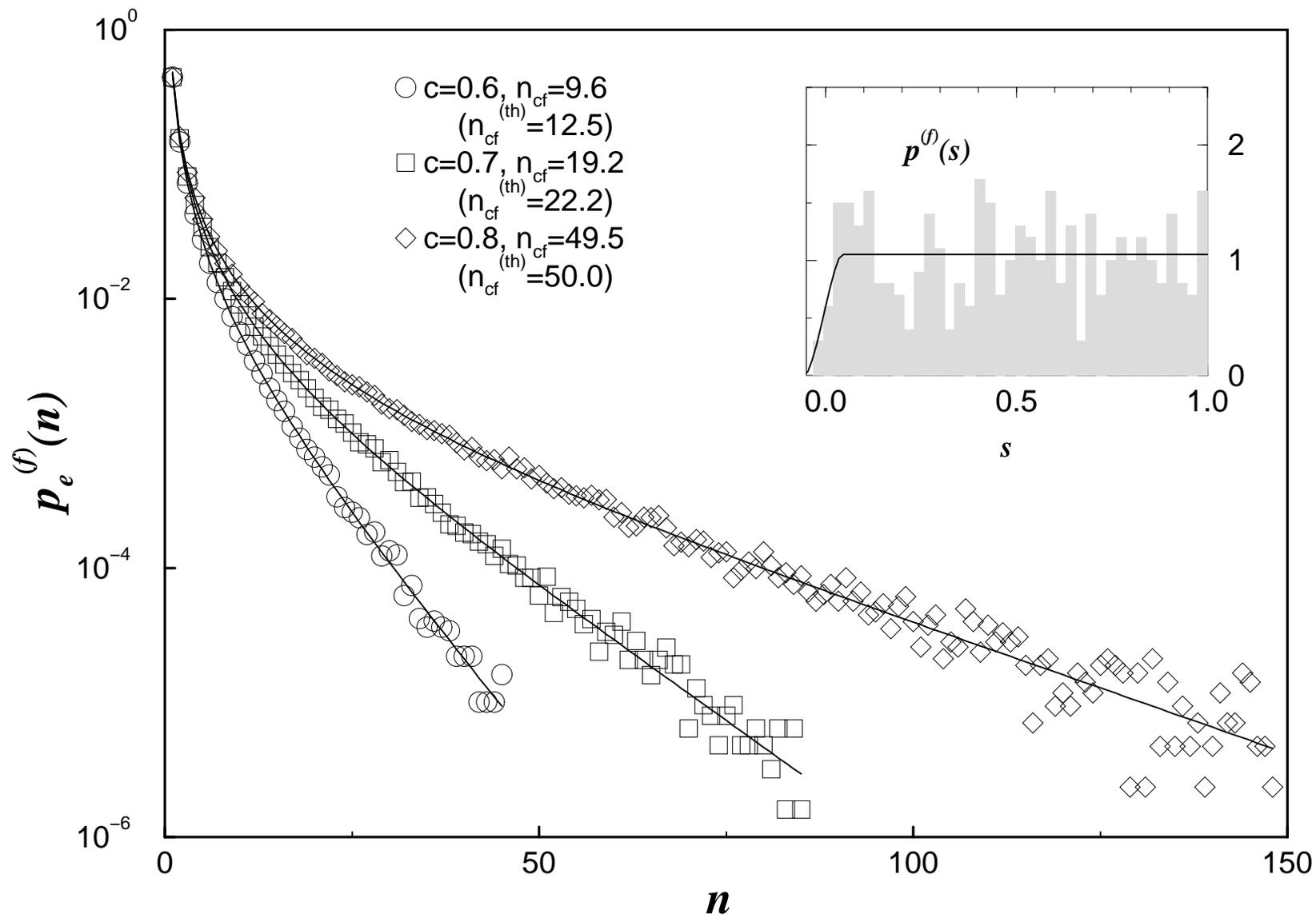
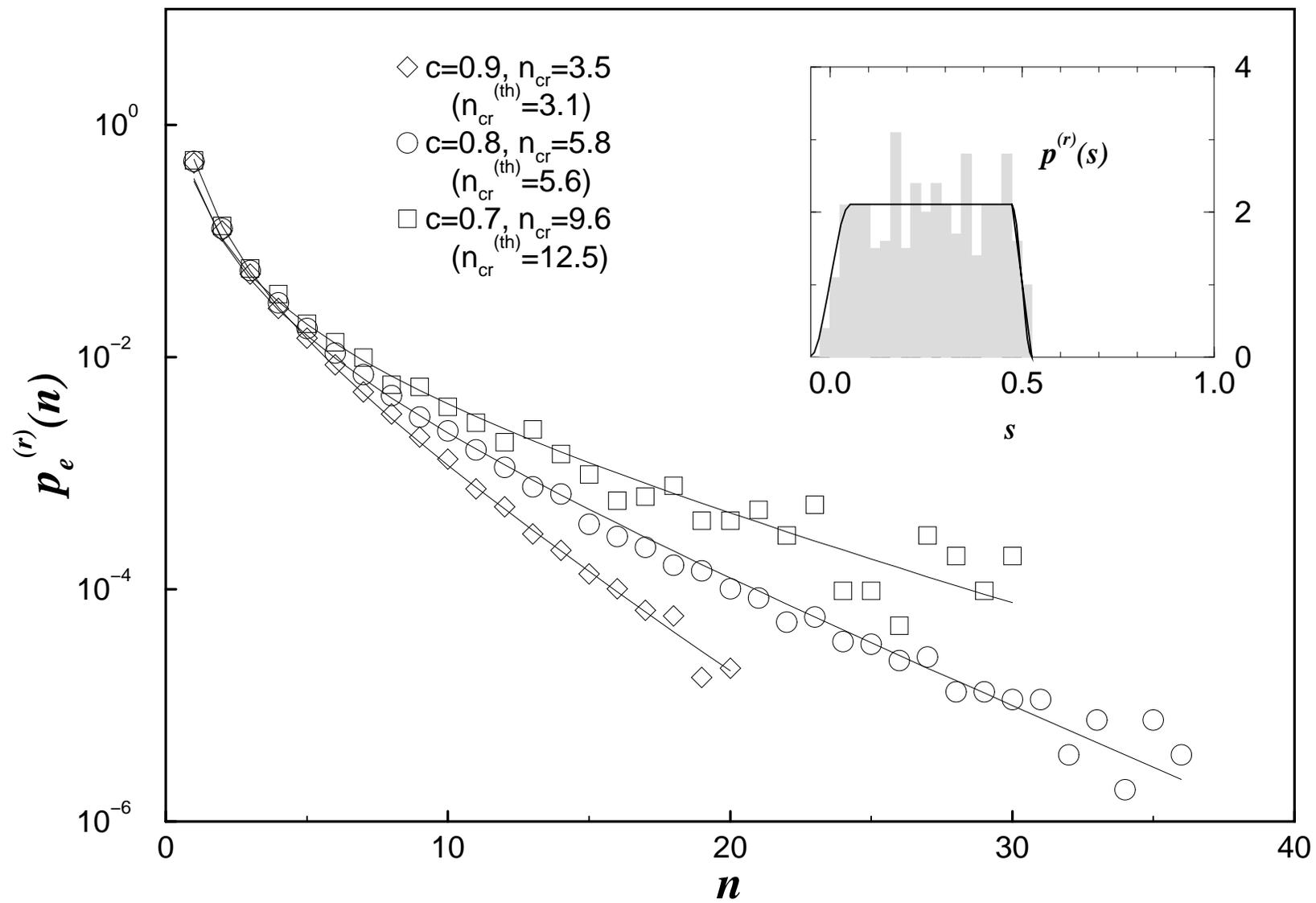
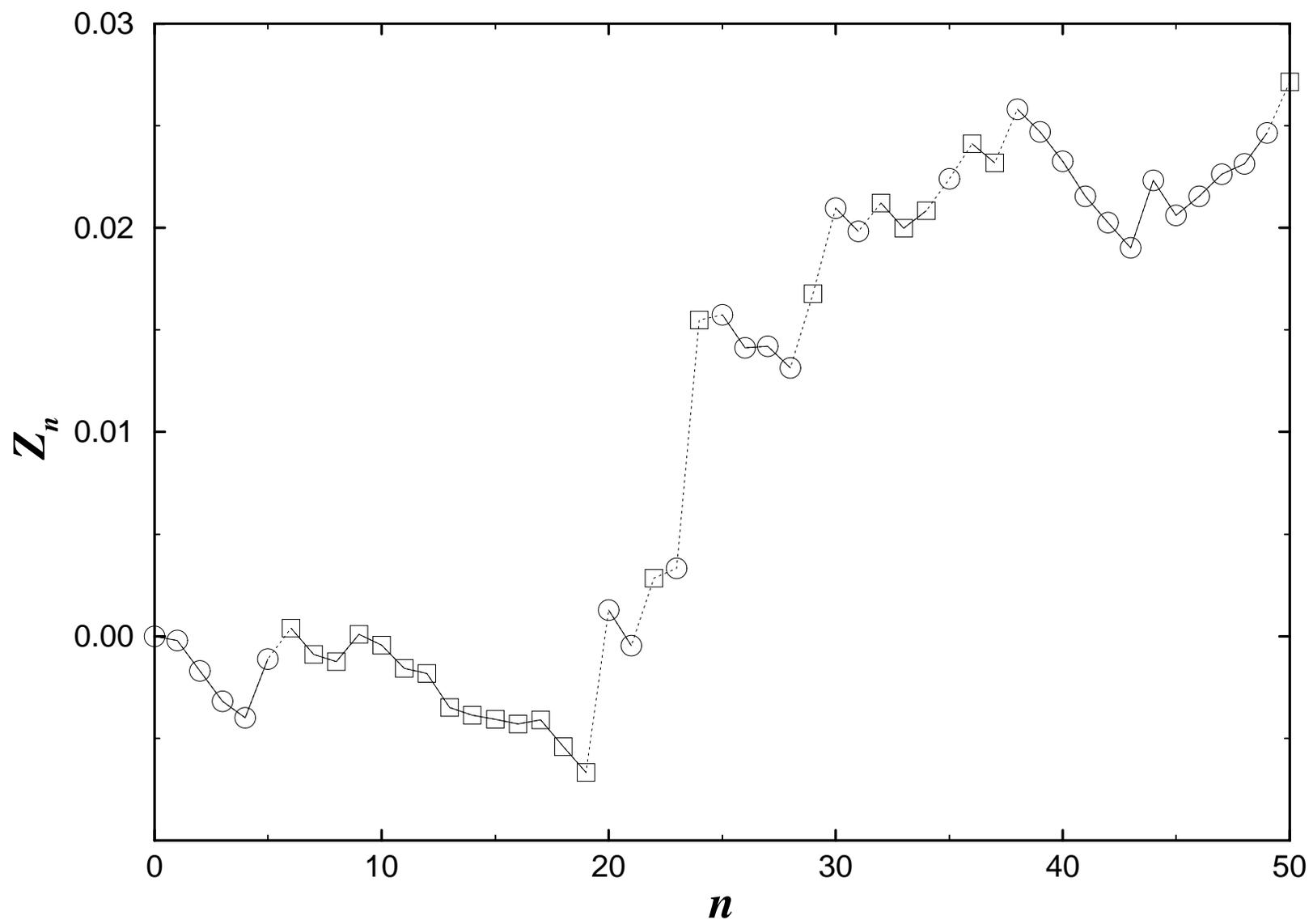
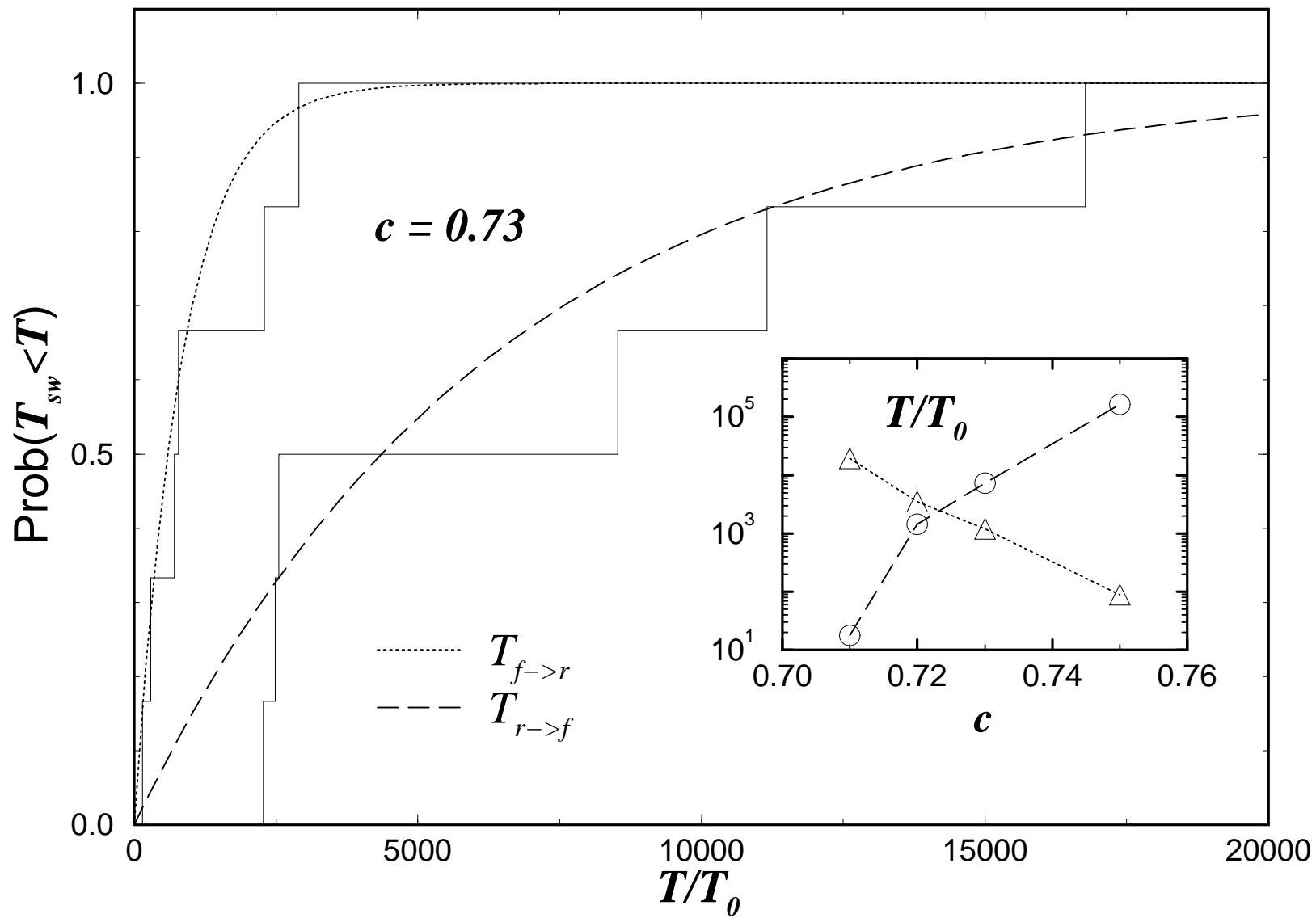
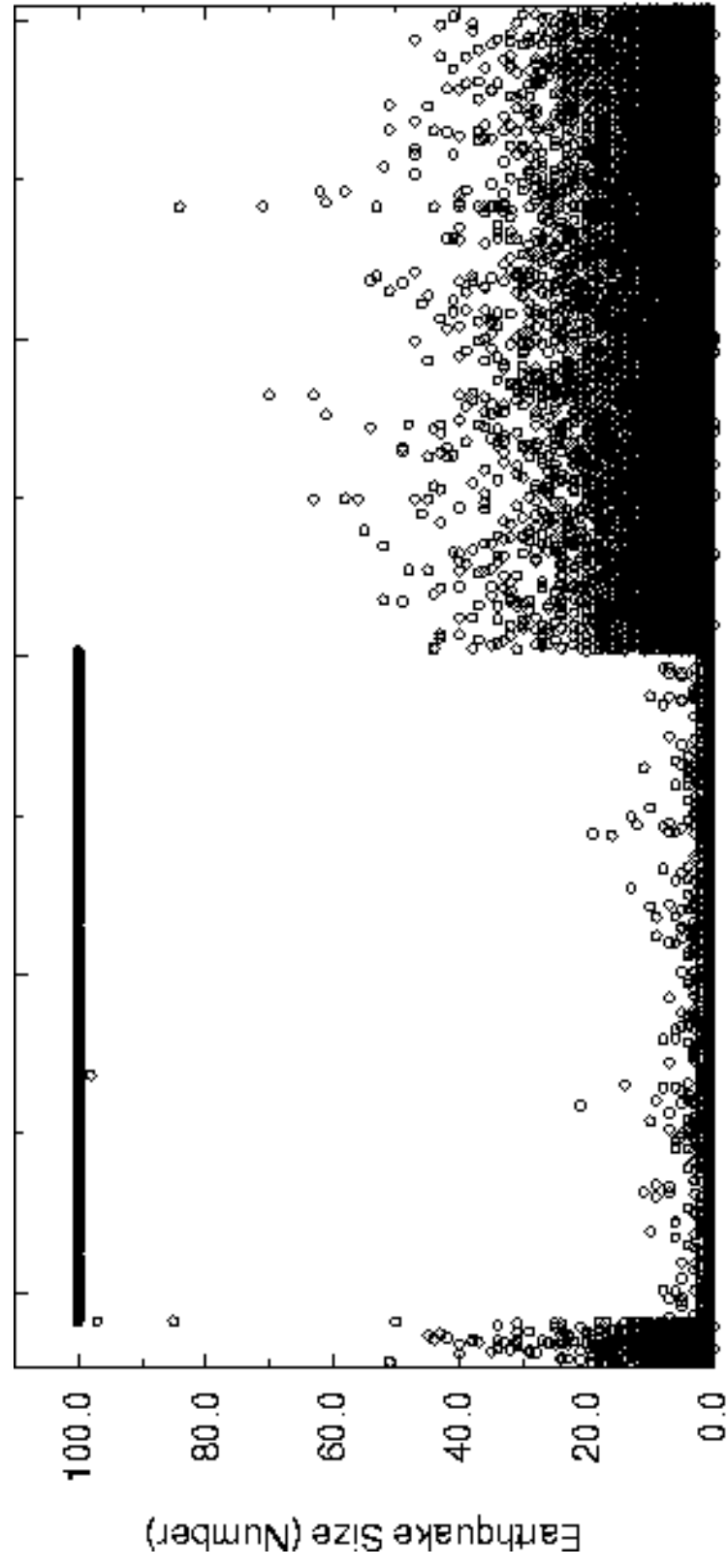
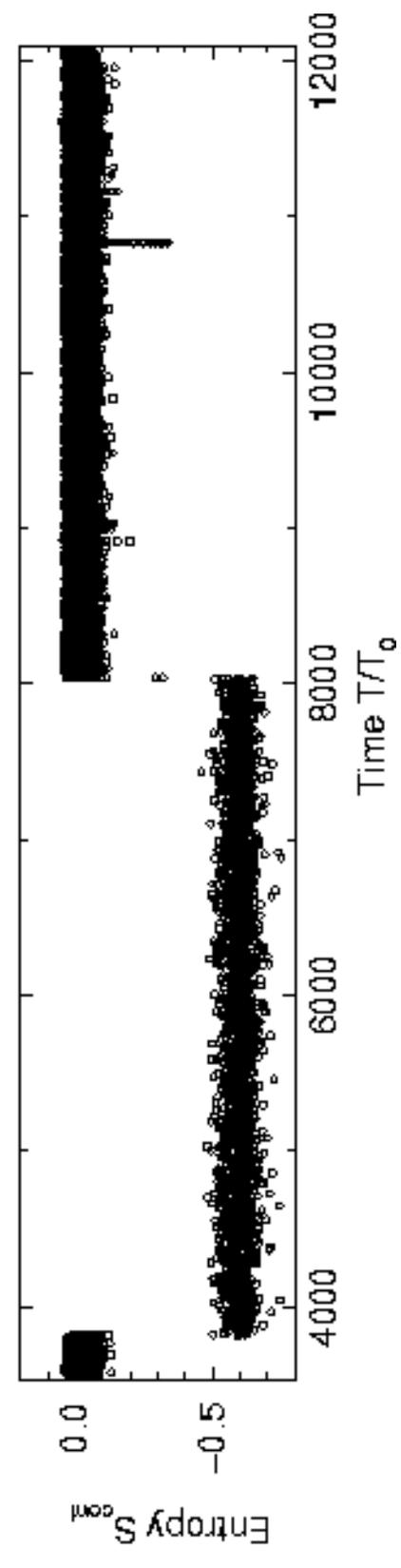
\end{document}